\documentclass[12pt,preprint]{aastex}
\usepackage{graphicx}

\def\gsim { \lower .75ex \hbox{$\sim$} \llap{\raise .27ex \hbox{$>$}} }
\def\lsim { \lower .75ex \hbox{$\sim$} \llap{\raise .27ex \hbox{$<$}} }
\def\be{\begin{equation}}
\def\ee{\end{equation}}
\def\bea{\begin{eqnarray}}
\def\eea{\end{eqnarray}}

\newcommand{\npb}{{Nucl.~Phys.~B}}
\newcommand{\plb}{{Phys.~Lett.~B}}
\newcommand{\ijmpd}{{Int.~J.~Mod.~Phys.~D}}

\shorttitle{On using the WMAP distance information in constraining
the time evolving equation of state of dark energy} \shortauthors{Li
{\it et al.}}

\begin{document}

\title{On using the WMAP distance information in constraining\\
the time evolving equation of state of dark energy}

\author{Hong Li\altaffilmark{1}, Jun-Qing Xia\altaffilmark{2},
Gong-Bo Zhao\altaffilmark{3}, Zu-Hui Fan\altaffilmark{1} \& Xinmin
Zhang\altaffilmark{2}}

\altaffiltext{1}{Department of Astronomy, School of Physics, Peking
University, Beijing, 100871, P. R. China.}

\altaffiltext{2}{Institute of High Energy Physics, Chinese Academy
of Science, P. O. Box 918-4, Beijing 100049, P. R. China;
xiajq@mail.ihep.ac.cn.}

\altaffiltext{3}{Department of Physics, Simon Fraser University,
Burnaby, BC, V5A 1S6, Canada.}

\begin{abstract}
Recently, the WMAP group has published their five-year data and
considered the constraints on the time evolving equation of state of
dark energy for the first time from the WMAP distance information.
In this paper, we study the effectiveness of the usage of these
distance information and find that these compressed CMB information
can give similar constraints on dark energy parameters compared with
the full CMB power spectrum if dark energy perturbations are
included, however, once incorrectly neglecting the dark energy
perturbations, the difference of the results are sizable.

\end{abstract}



\maketitle

\section{Introduction}

The newly released Wilkinson Microwave Anisotropy Probe five-year
data (WMAP5) \cite{Hinshaw08,Nolta08,Dunkley08,Komatsu}, detecting
the Cosmic Microwave Background (CMB) to an unprecedented precision,
make it possible to improve the constraints on almost all the
cosmological parameters, including the equation-of-state (EoS) $w$
of the unknown energy budget, dark energy. Defined as the ratio of
pressure over energy density, $w=P/\rho$, EoS can be used to
classify various dark energy models, such as quintessence
\cite{quint4,quint1,quint2,quint3}, phantom \cite{phantom}, quintom
\cite{quintom}, k-essence\\ \cite{kessence1,kessence2}, \emph{etc.},
which is of great theoretical significance to unveil the mystery of
dark energy. Therefore, trying to study the evolution history of EoS
of dark energy plays a crucial role in modern observational
cosmology \cite{recons3,recons2,recons1}. Simply put, one can choose
an arbitrary parametrization of $w$ and constrain the introduced
dark energy parameters from the astronomical observational data,
including CMB, Supernova type Ia (SN Ia), Large Scale Structure
(LSS) and so forth \cite{glof3,glof4,glof2,glof1}.

Recently, the WMAP group has released their five-year data and for
the first time considered the constraints on the time evolving EoS
of dark energy using the WMAP distance information. This method has
the advantage of reducing computation time by orders of magnitude,
yet the effectiveness and the level of approximation compared with
the full CMB power spectrum computation remain unclear. In this
paper, we make a thorough test of this simplified method to
investigate whether it is safe to constrain dark energy with time
evolving EoS. Our paper is structured as follows: In Section II we
describe the method and the data; in Section III we present our main
results; finally we present our conclusions in Section IV.

\section{Method and data}

To study the dynamical behavior of dark energy, we choose the
parametrization of the time evolving EoS of dark energy given by
\cite{linp1,linp2,Komatsu}:
\begin{equation}
\label{EoS} w(a) = w_{0} + w_{a}(1-a)~,
\end{equation}
where $a=1/(1+z)$ is the scale factor and $w_{a}=-dw/da$
characterizes the ``running" of the EoS (RunW henceforth). For the
$\Lambda$CDM model, $w_0=-1$ and $w_a=0$.

When using the MCMC global fitting strategy to constrain
cosmological parameters, it is crucial to include dark energy
perturbations, especially for the time evolving EoS of dark energy
models. This issue has been realized by many researchers including
the WMAP group
\cite{pertother2,pertother1,Zhao:2005vj,globf05,wmap3}. However,
when the parameterized EoS crosses $-1$, one cannot handle the dark
energy perturbations based on quintessence, phantom, k-essence and
other non-crossing models. By virtue of quintom, the perturbations
at the crossing points are continuous, thus we have proposed a
technique to treat dark energy perturbations in the whole parameter
space. For details of this method, we refer the readers to our
previous papers \cite{Zhao:2005vj,globf05}.

In this study, we have modified the publicly available Markov
Chain Monte Carlo package
CosmoMC\footnote{http://cosmologist.info/cosmomc/.} \cite{CosmoMC}
to include the dark energy perturbations with EoS across $-1$.
Furthermore, we assume purely adiabatic initial conditions and a
flat universe. Our most general parameter space is:
\begin{equation}
\label{para1} {\bf P} \equiv \left(\omega_{b}, \omega_{c},
\Theta_{s}, \tau,  w_{0}, w_{a}, n_{s}, \ln(10^{10}A_{s})\right)~,
\end{equation}
where $\omega_{b}\equiv\Omega_{b}h^{2}$ and
$\omega_{c}\equiv\Omega_{c}h^{2}$, where $\Omega_{b}$ and
$\Omega_{c}$ are the baryon and cold dark matter densities relative
to the critical density, $\Theta_{s}$ is the ratio (multiplied by
100) of the sound horizon at decoupling to the angular diameter
distance to the last scattering surface, $\tau$ is the optical depth
to reionization, $A_{s}$ and $n_{s}$ are the amplitude and the tilt
of the power spectrum of primordial scalar perturbations. For the
pivot scale of the primordial spectrum we set $k_{\ast} = 0.05$
Mpc$^{-1}$.

The WMAP distance information used by WMAP group include the ``shift
parameter" $R$, the ``acoustic scale" $l_{A}$ and the photon
decoupling epoch $z_{\ast}$. $R$ and $l_A$ correspond to the ratio
of angular diameter distance to the decoupling era over Hubble
horizon and sound horizon at decoupling respectively, given by
\begin{eqnarray}
R(z_{\ast})&=&\sqrt{{\Omega_m}H_0^2}\chi(z_{\ast})~,\\
l_{A}(z_{\ast})&=&\pi\chi(z_{\ast})/\chi_{s}(z_{\ast})~,
\end{eqnarray}
where $\chi(z_{\ast})$ and $\chi_s(z_{\ast})$ denote the comoving
distance to $z_{\ast}$ and comoving sound horizon at $z_{\ast}$
respectively. The decoupling epoch $z_{\ast}$ is given by \cite{Hu}
\begin{equation}
\label{defzstar} z_{\ast}=1048[1+0.00124(\Omega_b
h^2)^{-0.738}][1+g_1(\Omega_m h^2)^{g_2}]~,
\end{equation}
where \begin{equation} g_1=\frac{0.0783(\Omega_b
h^2)^{-0.238}}{1+39.5(\Omega_b h^2)^{0.763}},~
g_2=\frac{0.560}{1+21.1(\Omega_b h^2)^{1.81}}.
\end{equation}

The WMAP distance information encode in part of the CMB information
and can constrain cosmological parameters to some extent. It is
worth carefully investigating on the effectiveness of the
constraints from the distance information compared with the full CMB
power spectrum computation. To do this, we follow the procedure
shown in the flow chart:

\begin{eqnarray}\label{diag}
&{\rm Full~WMAP5~Data} &~ \stackrel{\sf MCMC}{\Huge \rightarrow}~ w_0,w_a \nonumber\\
{\sf MCMC}\hspace{-18mm}&\downarrow&~~~~~~~~~~~~\Updownarrow ~{\tt Compare}\nonumber\\
&l_A, R, z_{\ast}&~\stackrel{\sf MCMC}{\rightarrow}\hspace{3mm}
w_0,w_a
\end{eqnarray}

which are detailed as follows:
\begin{enumerate}
    \item Making a global fitting with MCMC method to constrain $w_{0}$,
$w_{a}$, and also $l_A$, $R$ and $z_{\ast}$ using the full WMAP5
power spectrum. In this step, we have done two types of
calculations, one with and the other without dark energy
perturbations;
    \item Using the
resultant $l_A$, $R$ and $z_{\ast}$ to constrain dark energy
parameters $w_{0}$ and $w_{a}$;
    \item Comparing the results of the constraints on $w_0$ and $w_a$
    obtained from step 1 and 2.
\end{enumerate}

In step 1, we calculate the likelihood of CMB power spectrum using
the routine supplied by the WMAP group\footnote{Legacy Archive for
Microwave Background Data Analysis (LAMBDA),
http://lambda.gsfc.nasa.gov/.}. In step 2, we calculate the
likelihood of WMAP distance information as follows \cite{Komatsu}:
\begin{equation}\label{priorlike}
\chi^2\equiv-2\ln{L}=(x^{th}_i-x^{data}_i)(C^{-1})_{ij}(x^{th}_j-x^{data}_j)~,
\end{equation}
where $x=(l_A,R,z_{\ast})$ is the parameter vector and
$(C^{-1})_{ij}$ is the inverse covariance matrix for the WMAP
distance information.

Since the purpose of this paper is not to make a global analysis,
in order to see the effects of the other cosmological data, we
include the gold sample of 182 SN Ia \cite{riess182} for a joint
constraint on EoS of dark energy in the combination with the WMAP5
data. In this study we also make use of Hubble Space Telescope
(HST) measurement of the Hubble parameter $H_{0}\equiv
100h$~km~s$^{-1}$~Mpc$^{-1}$ by multiplying the likelihood by a
Gaussian likelihood function centered around $h=0.72$ and with a
standard deviation $\sigma=0.08$ \cite{Hubble}, and a Gaussian
prior on the baryon density $\Omega_{b}h^{2}=0.022\pm0.002$
($1\sigma$) from Big Bang Nucleosynthesis (BBN) \cite{BBN}.

\section{Results}

\begin{figure}[htbp]
\begin{center}
\includegraphics[scale=0.52]{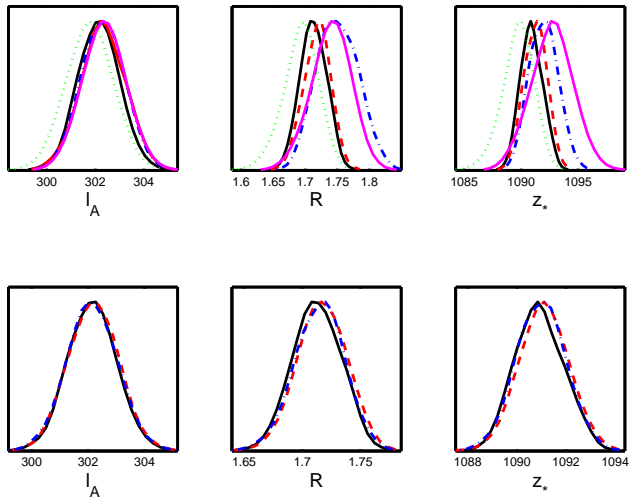}
\caption{One dimensional posterior distributions of $l_A$, $R$ and
$z_{*}$ with the WMAP5 data for different cosmological models. In
the upper panels, the black solid line is given by the standard flat
$\Lambda$CDM model, while the red dashed line, the blue dash-dotted
line, purple solid line and the green dotted line are given by
$\Lambda$CDM with non-zero $\Omega_k$, flat $\Lambda$CDM with
$f_{\nu}$, with $\alpha_s$ and with $r$ respectively. In the lower
panels, the black solid line is still from the standard $\Lambda$CDM
model, while the red dashed line and the blue dash-dotted lines are
given by the dark energy model with time evolving EoS (RunW model)
with and (incorrectly) without dark energy perturbations
respectively. \label{pc}}
\end{center}
\end{figure}

The WMAP distance information are extracted from the full WMAP5
power spectrum by assuming a certain cosmological model, and they
should be model dependent. In Fig.\ref{pc} we present the one
dimensional distributions of the WMAP distance information for
different cosmological models.

In the upper three panels of Fig.\ref{pc} we show the distributions
of $l_A$, $R$ and $z_{\ast}$ for five cosmological models: flat
$\Lambda$CDM model; $\Lambda$CDM with curvature; flat $\Lambda$CDM
model with massive neutrinos, with running of spectral index and
with tensor perturbations, respectively. We find that the
distributions of $R$ and $z_{\ast}$ are quite different in these
five cases, while the acoustic scale $l_{A}$ does not change
significantly. These results indicate that when using these distance
information to constrain cosmological parameters, one should be
clear about the assumed cosmological model. In Table I we also list
the median $1\sigma$ constraints on the WMAP distance information
from the full WMAP5 data for different cosmological models.

\begin{table}\label{table1}{\footnotesize
Table I. Median $1\sigma$ constraints on WMAP distance information
using full WMAP5 data for different cosmological models.
\begin{center}
\begin{tabular}{c|c|c|c}
  \hline
  \hline
 Models & $l_A$ & $R$ & $z_{*}$ \\
  \hline
  $\Lambda$CDM & $302.15\pm0.842$ & $1.71\pm0.021$ & $1090.92\pm0.969$\\ \hline
  $\Lambda$CDM +~$\Omega_K$ & $302.32\pm0.899$ & $1.72\pm0.021$ & $1091.26\pm1.004$\\ \hline
  $\Lambda$CDM +$~m_{\nu}$ & $302.30\pm0.873$ & $1.75\pm0.031$ & $1091.98\pm1.244$ \\ \hline
  $\Lambda$CDM +$~\alpha_s$ & $302.36\pm0.878$ & $1.74\pm0.031$ & $1092.72\pm1.817$ \\ \hline
  $\Lambda$CDM +$~r$~~ & $301.76\pm0.944$ & $1.69\pm0.027$ & $1089.72\pm1.366$ \\ \hline
  RunW with pert. & $302.20\pm0.865$ & $1.72\pm0.021$ & $1091.10\pm0.991$ \\ \hline
  RunW w/o pert. & $302.14\pm0.875$ & $1.71\pm0.021$ & $1090.97\pm0.985$ \\
  \hline
  \hline
\end{tabular}
\end{center}}

\end{table}

In the lower three panels of Fig.\ref{pc} we show the results for
three flat models with different dark energy properties:
$\Lambda$CDM model, RunW model with and (incorrectly) without dark
energy perturbations. These results do not show significant
differences in the WMAP distance information among different dark
energy models. We also compare the results obtained with and
(incorrectly) without dark energy perturbations and find that simply
switching off dark energy perturbations does not bias the results
much at this stage\footnote{The distance information are determined
by the background parameters and not affected by dark energy
perturbations significantly.}. In the following calculations, we use
the WMAP distance information obtained from the RunW model with dark
energy perturbations included. The corresponding inverse covariance
matrix is shown in Table II.

\begin{table}\label{table2}
Table II. Inverse covariance matrix for the WMAP distance
information $l_A$, $R$ and $z_{\ast}$ in RunW dark energy model when
including dark energy perturbations.
\begin{center}
\begin{tabular}{cccc}
  \hline
  \hline
    &~~~$l_A(z_{\ast})$~~~&~~$R(z_{\ast})$~~~&~~~~~$z_{\ast}$~~~\\
  \hline
  ~~$l_A(z_{\ast})$ & $1.795$ & $31.596$ & ~~$-1.146$\\
  ~~$R(z_{\ast})$ &  & $5409.68$ & ~~$-94.58$\\
  ~~$z_{\ast}$ &  & & ~~$2.891$\\

  \hline
  \hline
\end{tabular}
\end{center}

\end{table}

\begin{figure}[htbp]
\begin{center}
\includegraphics[scale=0.45]{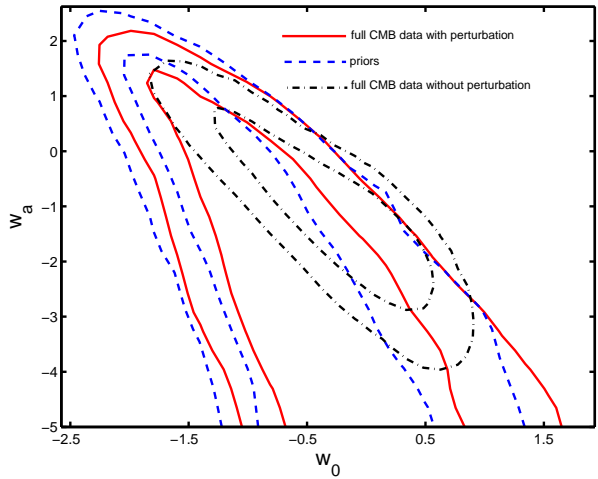}
\caption{$68\%$ and $95\%$ confidence levels constraints on
($w_0$,$w_a$) from full WMAP5 data and WMAP distance information
respectively. Red solid lines are obtained from the full WMAP5 data
including dark energy perturbations; black dash-dotted lines are
from the full WMAP5 data incorrectly neglecting dark energy
perturbations; and blue dashed lines are from WMAP distance
information. \label{w0w1cmb}}
\end{center}
\end{figure}

In Fig.\ref{w0w1cmb} we compare the constraints on $w_{0}$ and
$w_{a}$ obtained from the full WMAP5 power spectrum with the one
obtained from WMAP distance information given in Table I and II.
From this plot we can see that the WMAP distance information and the
full WMAP5 power spectrum with dark energy perturbations included
can give quite similar constraints on $w_{0}$ and $w_{a}$. However,
when the dark energy perturbations are incorrectly switched off
(black dash-dotted lines in Fig.\ref{w0w1cmb}), the results between
the two methods are quite different.

The WMAP distance information mainly include the information of the
oscillatory structures of the CMB power spectrum, which come from
the small angular scale (large $l$) of the power spectrum. On the
other hand, for the full CMB power spectrum, they combine more
information than the distance information, especially at large
angular scale (small $l$). At large angular scale, they are affected
by the late Integrated Sachs-Wolfe (ISW) effects, which are dark
energy dependent. Thus tighter constraints on ($w_0,w_a$) are
anticipated by using the full CMB spectrum than those from using the
distance priors only. This is clearly demonstrated in
Fig.\ref{w0w1cmb} (dashed contours versus dash dotted one). It is
noted that the dash dotted contours are calculated without including
dark energy perturbations, and thus the constraining power of the
late ISW effect on dark energy parameters is fully realized.
However, when including the dark energy perturbations, which are
mainly effective at small $l$, the constraints on dark energy
parameters from the late ISW effects are significantly reduced,
resulting in similar contours shown by the dashed and solid lines in
Fig.\ref{w0w1cmb} \footnote{In our analysis we use a specific
parametrization the ``RunW" model. We expect that our results hold
qualitatively for other dark energy parameterizations.
Quantitatively however, the specific results are dependent on the
detailed calculations on different dark energy parameterizations
that are used.} \cite{communication}. The differences between the
solid and the dash dotted contours also show that how biased results
can be obtained if the dark energy perturbations are incorrectly
neglected in the full CMB data analysis \cite{globf05,wmap3}.

\begin{figure}[htbp]
\begin{center}
\includegraphics[scale=0.45]{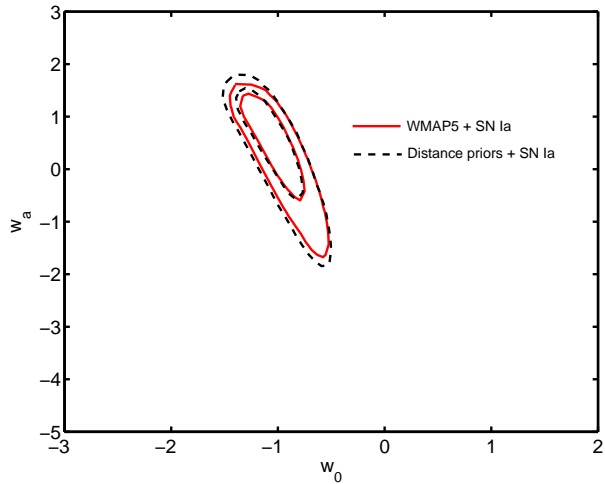}
\caption{Two-dimensional constraints on the parameters of dark
energy from the combined WMAP5 and SN Ia. The red solid and black
dotted lines are the constraints from ``full WMAP5 data + SN Ia"
with dark energy perturbations and ``WMAP distance information + SN
Ia" respectively. \label{w0w1cmbsn}}
\end{center}
\end{figure}

In Fig.\ref{w0w1cmbsn}, we give the constraints on dark energy
parameters by adding the SN Ia data. We can see that the constraints
on dark energy parameters are tightened and the differences between
the results obtained from ``full WMAP5 power spectrum + SN Ia" and
from ``WMAP distance information + SN Ia" become insignificant.

\section{Summary}
In this paper, we have studied the effectiveness of the WMAP
distance information on constraining the dark energy parameters, by
comparing with the full WMAP5 power spectrum analysis. We first
present the level of the model dependence of the distance
information in different cosmological models. We further clarify
that by taking into account dark energy perturbations properly, the
WMAP distances can give unbiased information on dark energy
parameters relative to the full CMB analysis.

\section*{Acknowledgments}

We acknowledge the use of the Legacy Archive for Microwave
Background Data Analysis (LAMBDA). Support for LAMBDA is provided
by the NASA Office of Space Science. We have performed our
numerical analysis in the Shanghai Supercomputer Center (SSC). We
would like to thank Eiichiro Komatsu, Tao-Tao Qiu and Hua-Hui
Xiong for discussions. This work is supported in part by the
funding support from the China postdoctoral science foundation and
the National Natural Science Foundation of China under Grant Nos.
90303004, 10533010 and 10675136 and by the Chinese Academy of
Science under Grant No. KJCX3-SYW-N2. GZ is supported by National
Science and Engineering Research Council of Canada (NSERC).

\end{document}